\documentclass[a4paper,11pt]{article}
\usepackage{pos}

\title{Range correction in the weak-binding relation for unstable states}

\author*{Tomona Kinugawa}
\author{Tetsuo Hyodo}

\affiliation{Department of Physics Tokyo Metropolitan University,\\
  Hachioji 192-0397, Japan}

\emailAdd{kinugawa-tomona@ed.tmu.ac.jp}
\emailAdd{hyodo@tmu.ac.jp}

\abstract{
The compositeness is defined as the weight of the hadronic molecule in the hadron wave function. We can determine the internal structure of the weak-binding system without any specific models from the compositeness. In order to estimate the compositeness of the system with a large effective range, we introduce the range correction to Weinberg's weak-binding relation by modifying the correction terms. We study the applicability of the weak-binding relations by the numerical calculation and show that the improved relation can be applied to a larger parameter region compared with the previous one.

}

\FullConference{%
  *** Particles and Nuclei International Conference - PANIC2021 ***\\
  *** 5 - 10 September, 2021 ***\\
  *** Online ***
}

\begin{document}
\maketitle

\section{Introduction}
Hadrons are the particles which interact with each other by the strong interaction. Almost all hadrons consist of two or tree quarks. The hadrons which cannot be classified as two or three quarks are called exotic hadrons. Many candidates for exotic hadrons are discovered in recent experiments~\cite{Hosaka:2016pey,Guo:2017jvc}. Those candidates are considered as, for example, multiquark states or hadronic molecule states. The multiquark states are regarded as the compact states of four quarks or more. The hadronic molecule states are considered as weakly bound states of hadrons.


In order to analyze the internal structure of the candidates for exotic hadrons, we focus on the compositeness $X$ which can be interpreted as the weight of the hadronic molecule component~\cite{Hyodo:2011qc}. If the system is a complete hadronic molecule state, $X=1$. When the binding energy is small, we can estimate the compositeness $X$ with the weak-binding relation for stable states~\cite{Weinberg:1965zz,Kamiya:2015aea,Kamiya:2016oao}:
\begin{align}
a_0=R\left\{\frac{2X}{1+X}+\mathcal{O}\left(\frac{R_{\rm typ}}{R}\right)\right\},
\label{eq:wbr_s}
\end{align}
where $a_{0}$ is the scattering length, $R\equiv1/\sqrt{2\mu B}$ is the length scale determined by the binding energy $B$ and the reduced mass $\mu$. $R_{\rm typ}$ is regarded as the interaction range. When the binding energy is small such that $R_{\rm typ}\ll R$, we can neglect the correction terms of the relation~\eqref{eq:wbr_s} and estimate $X$ only from the observables $a_{0}$ and $R$ without any specific models.


Let us consider the weak-binding relation~\eqref{eq:wbr_s} from the viewpoint of the low-energy universality~\cite{Braaten:2004rn,Naidon:2016dpf}. In the weak-binding limit $R\to\infty$, magnitude of the scattering length $|a_{0}|$ is very large and the system becomes scale invariant, namely all lengths in the system are scaled by $a_{0}$. In this limit, $a_{0}=R$, and the bound state is completely a hadronic molecule. When we are away from the weak-binding limit, $a_{0}$ deviates from $R$ and the deviations are expressed by the factor in the parenthesis in Eq.~\eqref{eq:wbr_s}. We can interpret $2X/(1+X)$ and $\mathcal{O}(R_{\rm typ}/R)$ terms in the weak-binding relation as the origins of the deviations, corresponding to contributions other than the hadronic molecule component $(X\neq 1)$ and the nonzero interaction range $(R_{\rm typ}\neq 0)$, respectively. In this work, we discuss the range correction in the weak-binding relation for systems with a large effective range $r_{e}$. Because the effective range is one of the length scales which characterize the low-energy scattering, it is expected that $r_{e}$ induces another deviation of $a_{0}$ from $R$ in addition to the above two.


\section{Range correction in the weak-binding relation}

To discuss the range correction, we apply the weak-binding relation to the single-channel models with a zero-range interaction $R_{\rm typ}\to 0$. In such models, because the system has no channel couplings, $X=1$ by the definition of $X$. Furthermore, with the zero-range interaction, the correction terms of the weak-binding relation $\mathcal{O}(R_{\rm typ}/R)$ are exactly zero. Therefore, it seems that these models always show $a_{0}=R$ according to the weak-binding relation.

 However, we find that the effective range model~\cite{Braaten:2007nq} is the exception of this case as explained below. The effective range model is the single-channel non-relativistic effective field theory with a derivative coupling. The Hamiltonian is given by 
\begin{align}
\mathcal{H}_{\rm int}=\frac{1}{4}\lambda_0(\psi^\dagger\psi)^2+\frac{1}{4}\rho_0\nabla(\psi^\dagger\psi)\cdot\nabla(\psi^\dagger\psi).
\end{align} 
The scattering amplitude at momentum $k$ in the zero-range limit (cutoff $\Lambda\to \infty$) can be written by 
\begin{align}
f(k)=\left[-\frac{1}{a_0}+\frac{r_e}{2}k^2-ik\right]^{-1}. 
\end{align}
The eigenmomentum of the bound states is given by the pole of the scattering amplitude $|f(k)|\to \infty$. With the relation of $R=i/k$ at the pole, we can obtain the relation between $a_{0}$ and $R$:
\begin{align}
a_0=R\frac{2r_e/R}{1-(r_e/R-1)^2}=R\left\{1+\mathcal{O}\left(\left|\frac{r_e}{R}\right|\right)\right\}.
\end{align}
This result shows that $a_{0}$ deviates from $R$ in the effective range model because of $r_{e}\neq 0$ even in the single channel system with the zero-range interaction. Therefore, we conclude that the relation \eqref{eq:wbr_s} dose not hold in this case.


From the discussion of the effective range model in the zero-range limit, we propose the range correction in the weak-binding relation by modifying the correction terms. We denote the interaction range as $R_{\rm int}$ which  was previously denoted as $R_{\rm typ}$, and redefine $R_{\rm typ}$ as follows:
\begin{align}
\label{eq:range}
R_{\rm typ}&=\max\{R_{\rm int},R_{\rm eff}\},
\end{align}
where, $R_{\rm eff}$ is the maximum length scale in the effective range expansion except for $a_{0}$. With this new definition of $R_{\rm typ}$, there is no contradiction even in the effective range model with $R_{\rm int}=0$ and $R_{\rm eff}=|r_{e}|$. If $R_{\rm eff}\to0$, then $R_{\rm typ}=R_{\rm int}$ and the weak-binding relation reduce to the previous one. 


\section{Numerical calculations}
In this section, we perform the numerical calculations to study the applicable parameter region of the weak-dinging relations with comparing the redefined $R_{\rm typ}$ in Eq.~\eqref{eq:range} and the previous one, $R_{\rm typ}=R_{\rm int}$. When the observables $a_{0}$ and $R$ are given, we estimate the compositeness $X$ using the weak-binding relation~\eqref{eq:wbr_s}. The central value $X_{c}$ is given by the relation~\eqref{eq:wbr_s} neglecting the correction terms. For the estimation of the uncertainty of $X$, we define $\xi\equiv R_{\rm typ}/R$ according to Ref.~\cite{Kamiya:2016oao}. We determine the upper and lower boundaries $X_{u,l}$ with taking into account the leading order in $\xi$. In summary, $X_{c}$ and $X_{u,l}$ are written as
\begin{align}
X_c=\frac{a_0/R}{2-a_0/R},\quad X_{u,l}=\frac{a_0/R}{2-a_0/R}\pm\xi.
\end{align}


We study the validity of the estimation of $X$ from the weak-binding relation by using the model in which the exact value $X_{\rm exact}$ is known. For the estimation to be successful, we require the following condition:
\begin{itemize}
\item Validity condition : $X_{\rm exact}$ is contained in the uncertainty region, $X_l<X_{\rm exact}<X_u$.
\end{itemize}



We consider the effective range model with the finite interaction range $R_{\rm int}\neq 0$ $(\Lambda<\infty)$ for the numerical calculation. In this case, the scattering amplitude has two length scales $r_e$ and $R_{\rm int}$ except for $a_{0}$. Because both of those can be $R_{\rm typ}$, we need to calculate two cases, $R_{\rm typ}=|r_e|$ or $R_{\rm typ}=R_{\rm int}$ in the improved relation. The validity condition can be verified because the exact value $X_{\rm exact}$ is known to be 1 in this single channel model. We perform the numerical calculations of the uncertainty estimated by $r_{e}$ and $R_{\rm int}$, and search for the parameter region where the weak-binging relation can be applied.

\begin{figure}[htbp]
\centering
\includegraphics[width=0.45\textwidth,bb=59 44 786 505]{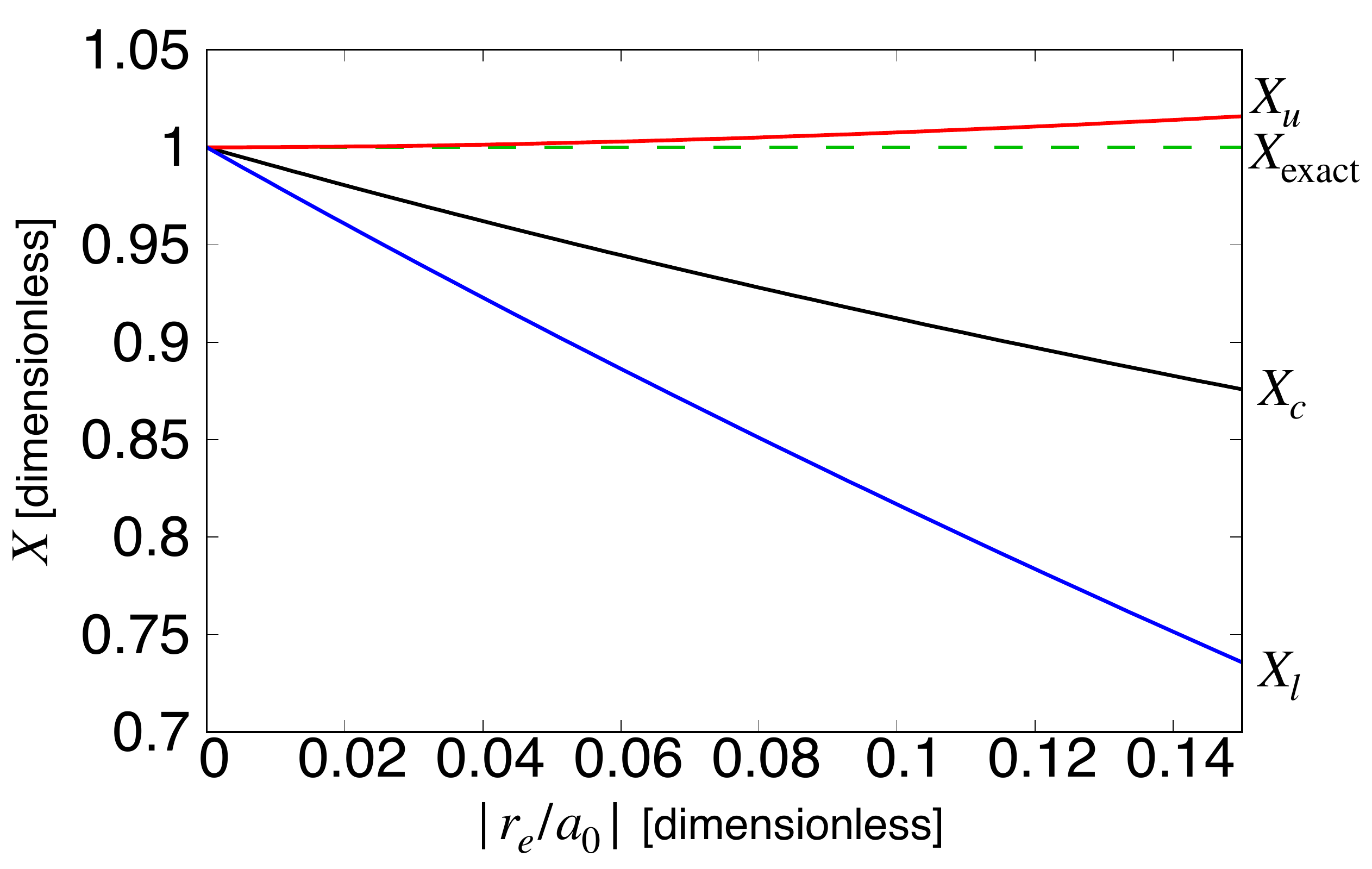}
\includegraphics[width=0.45\textwidth,bb=55 45 783 502]{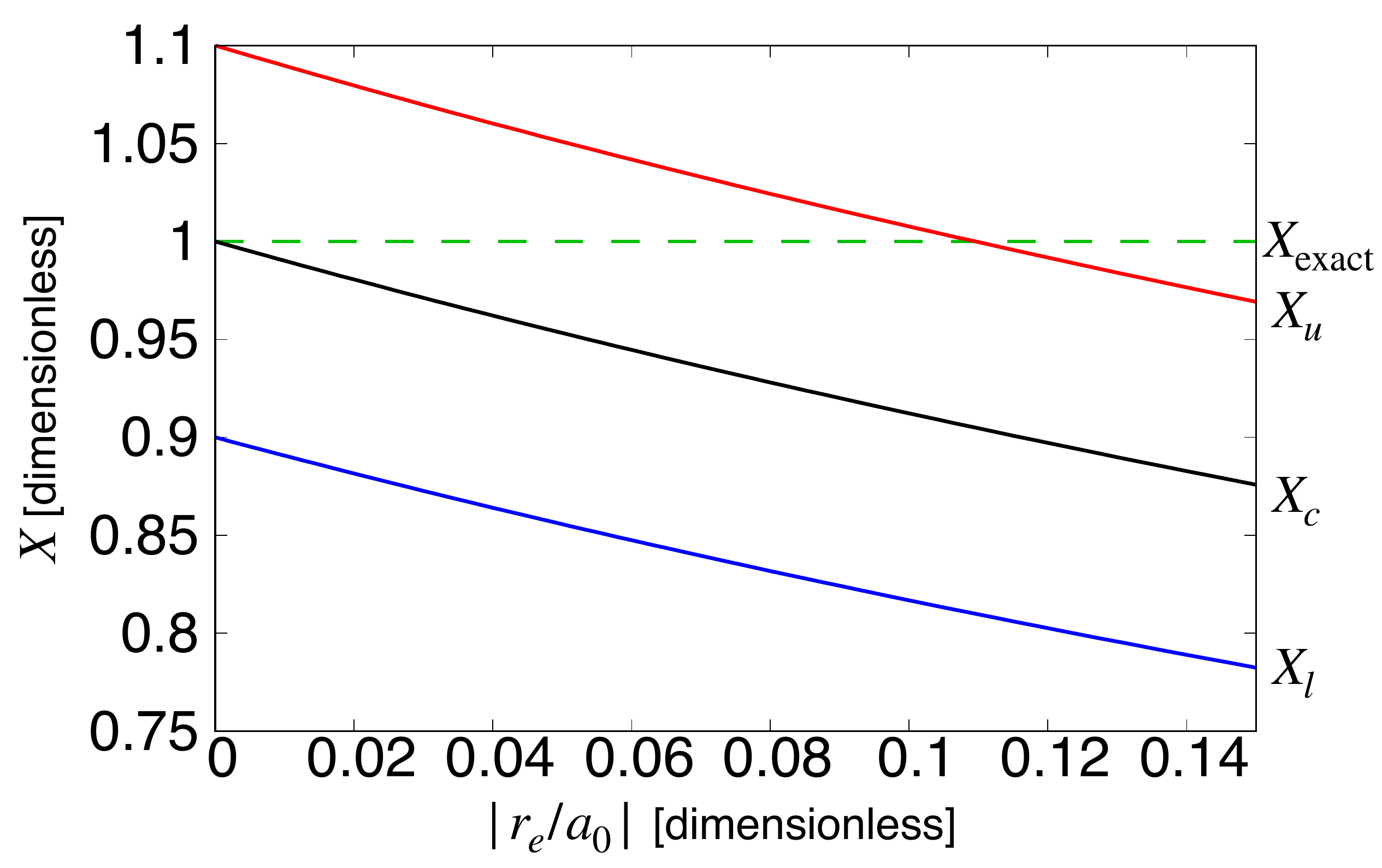}
\caption{The central value $X_{c}$ and the upper and lower boundaries $X_{u,l}$ by $R_{\rm typ}=|r_{e}|$ (left) and by $R_{\rm typ}=R_{\rm int}$ (right) at fixed $R_{\rm int}=0.1a_{0}$.}
\label{fig:X_boundary}
\end{figure}

In Fig.~\ref{fig:X_boundary}, we show the central value of $X$ together with the upper and lower boundaries at fixed $R_{\rm int}=0.1a_{0}$. The left (right) panel shows the uncertainty by $R_{\rm typ}=|r_{e}|$ ($R_{\rm typ}=R_{\rm int}$). Because we consider the $r_{e}<0$ case, we plot $X$ as a function of $|r_{e}/a_{0}|$. From the left panel, because $X_{\rm exact}=1$ is contained in the uncertainty region by $R_{\rm typ}=|r_{e}|$, the validity condition is satisfied in this parameter region. In contrast, right panel shows that $X_{\rm exact}$ is not contained in the uncertainty region by $R_{\rm typ}=R_{\rm int}$ for $|r_{e}/a_{0}|\gtrsim 0.1$. Therefore, the validity condition by $R_{\rm typ}=R_{\rm int}$ is not satisfied for the systems with a large $|r_{e}|$.
 

\begin{figure}[tbp]
\centering
\includegraphics[width=0.65\textwidth,bb=41 65 734 474]{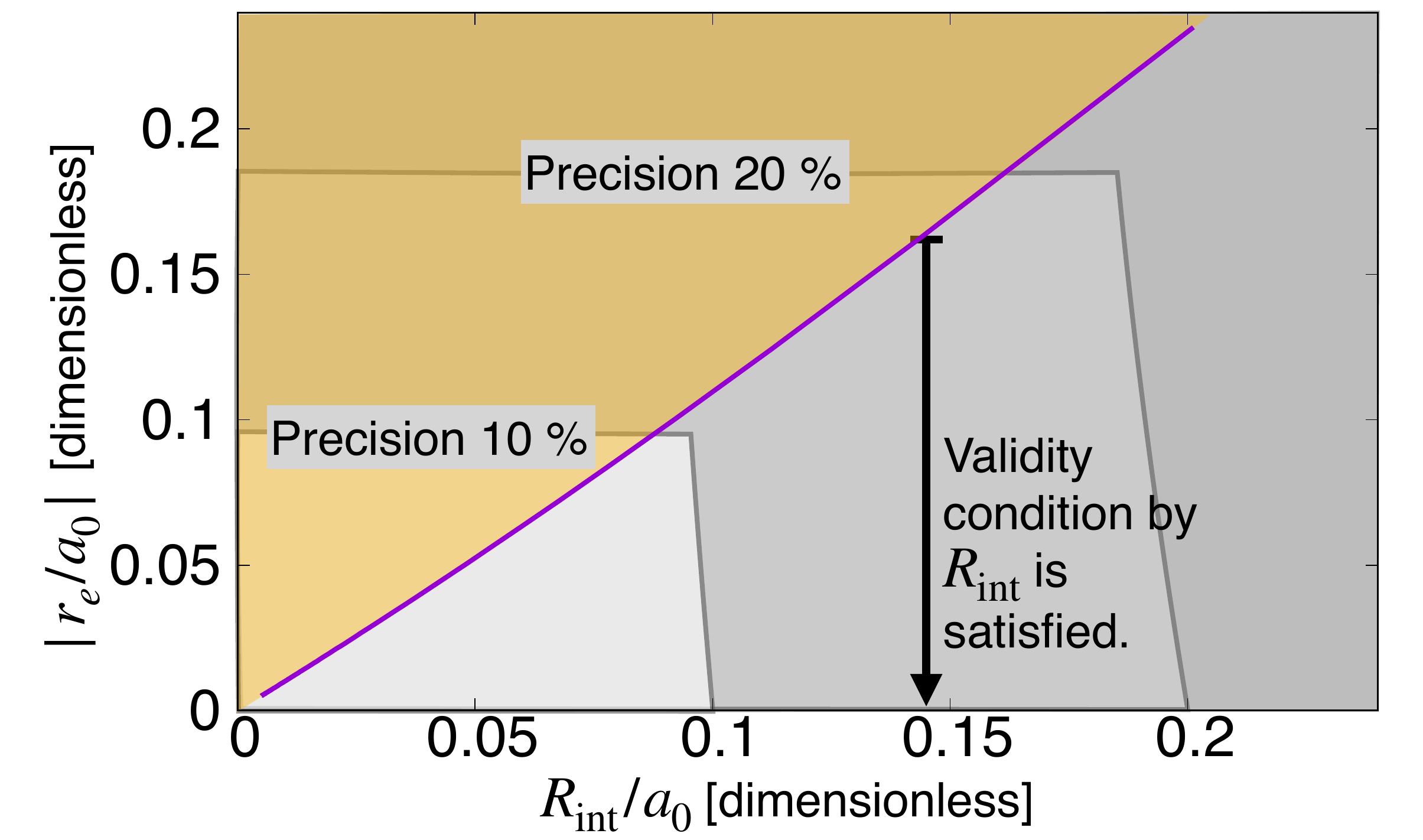}
\vspace{0.7cm}
\caption{Applicable regions of the previous and improved weak-binding relations in the $|r_e/a_0|-R_{\rm int}/a_0$ plane.}
\label{fig:region_plot}
\end{figure}

Performing the same calculations with varying $R_{\rm int}$, we plot the applicable parameter region of the previous and improved weak-binding relations in the $|r_e/a_0|-R_{\rm int}/a_0$ plane in Fig.~\ref{fig:region_plot}. The validity condition by $R_{\rm typ}=|r_{e}|$ is satisfied in the whole region of this plot because $X_{\rm exact}$ is always contained in the uncertainty region by $R_{\rm typ}=|r_{e}|$. However, the validity condition is only satisfied below the solid line when we consider the uncertainty by $R_{\rm typ}=R_{\rm int}$. From these numerical results, both the previous and improved weak-binding relations can be applied below the solid line of Fig.~\ref{fig:region_plot} because the validity conditions are satisfied both by $R_{\rm typ}=|r_{e}|$ and $R_{\rm typ}=R_{\rm int}$. In the region above the solid line, however, only the improved weak-binding relation can be applied because the uncertainty by $R_{\rm typ}=|r_{e}|$ is not considered in the previous relation. We conclude that our range correction works well because the applicable region of the improved weak-binding relation is larger than of the previous one.

In order to discuss the precision of the estimation of $X$, we define $P=|(X_{c}-X_{u,l})/X_{c}|$ as the magnitude of the normalized uncertainty. We show contour plot of the precision $P$ in Fig.~\ref{fig:region_plot}. For the meaningful estimation, $a_{0},r_{e}$ and $R_{\rm int}$ of the bound state should be in the small $P$ region.  This plot will be useful when we apply the weak-binding relation to the actual hadron systems.


\section{Summary and future prospects}
In this work, we study the weak-binding relation with which the compositeness of the weakly bound state can be estimated by the observables. We discuss the range correction for the system which has large effective range and propose the improvement of the weak-binding relation by redefining the correction terms. From the results of the numerical calculations, the improved weak-binding relation can be used in a larger parameter region than the previous one.

As the future prospect, we will discuss the range correction in the weak-binding relation for the unstable states with a finite decay width. As preliminary results, we check that the unstable state in the effective range model in the zero range limit has a similar wave function as the stable one. As the next step, we will study the applicability of the weak-binding relation with range correction for unstable states by the numerical calculations. The final goal is to apply the improved weak-binding relation for the actual hadron systems.




\begin{thebibliography}{99}
\bibitem{Hosaka:2016pey}
A.~Hosaka, T.~Iijima, K.~Miyabayashi, Y.~Sakai and S.~Yasui,
PTEP \textbf{2016}, no.6, 062C01 (2016)
[arXiv:1603.09229 [hep-ph]].

\bibitem{Guo:2017jvc}
F.~K.~Guo, C.~Hanhart, U.~G.~Mei\ss{}ner, Q.~Wang, Q.~Zhao and B.~S.~Zou,
Rev. Mod. Phys. \textbf{90}, no.1, 015004 (2018)
[arXiv:1705.00141 [hep-ph]].

\bibitem{Hyodo:2011qc}
T.~Hyodo, D.~Jido and A.~Hosaka,
Phys. Rev. C \textbf{85}, 015201 (2012)
[arXiv:1108.5524 [nucl-th]].

\bibitem{Weinberg:1965zz}
S.~Weinberg,
Phys. Rev. \textbf{137}, B672-B678 (1965).

\bibitem{Kamiya:2015aea}
Y.~Kamiya and T.~Hyodo,
Phys. Rev. C \textbf{93}, no.3, 035203 (2016)
[arXiv:1509.00146 [hep-ph]].

\bibitem{Kamiya:2016oao}
Y.~Kamiya and T.~Hyodo,
PTEP \textbf{2017}, no.2, 023D02 (2017)
[arXiv:1607.01899 [hep-ph]].

\bibitem{Braaten:2004rn}
E.~Braaten and H.~W.~Hammer,
Phys. Rept. \textbf{428}, 259-390 (2006)
[arXiv:cond-mat/0410417 [cond-mat]].

\bibitem{Naidon:2016dpf}
P.~Naidon and S.~Endo,
Rept. Prog. Phys. \textbf{80}, no.5, 056001 (2017)
[arXiv:1610.09805 [quant-ph]].

\bibitem{Braaten:2007nq}
E.~Braaten, M.~Kusunoki and D.~Zhang,
Annals Phys. \textbf{323}, 1770-1815 (2008)
[arXiv:0709.0499 [cond-mat.other]].

\end{thebibliography}
\end{document}